\documentclass[aps,prb,twocolumn,longbibliography,preprintnumbers,amsmath,amssymb,superscriptaddress]{revtex4-1}

\usepackage{graphicx}
\usepackage[T1]{fontenc}
\usepackage[colorlinks,bookmarks=false,citecolor=blue,linkcolor=red,urlcolor=blue]{hyperref}
\usepackage[dvipsnames]{xcolor}
\usepackage{times}

\usepackage{bbm}

\newcommand{\be}{\begin{eqnarray}}
\newcommand{\ee}{\end{eqnarray}}

\newcommand{\nn}{\nonumber } 
\newcommand{\Eqref}[1]{Eq.~\eqref{#1}}

\begin{document}

%-------------------------------------------------------------------------------------
\author{Daniel D. Scherer}
\affiliation{Niels Bohr Institute, University of Copenhagen, Lyngbyvej 2, DK-2100 Copenhagen, Denmark}
%-------------------------------------------------------------------------------------

%-------------------------------------------------------------------------------------
\author{Brian M. Andersen}
\affiliation{Niels Bohr Institute, University of Copenhagen, Lyngbyvej 2, DK-2100 Copenhagen, Denmark}
%-------------------------------------------------------------------------------------

\title{Effects of spin-orbit coupling on the neutron spin resonance in iron-based superconductors}

\begin{abstract}

The so-called neutron spin resonance consists of a prominent enhancement of the magnetic response at a particular energy and momentum transfer upon entering the superconducting state of unconventional superconductors. In the case of iron-based superconductors, the neutron resonance has been extensively studied experimentally, and a peculiar spin-space anisotropy has been identified by polarized inelastic neutron scattering experiments. Here we perform a theoretical study of the energy- and spin-resolved magnetic susceptibility in the superconducting state with $ s_{+-} $-wave order parameter, relevant to iron-pnictide and iron-chalcogenide superconductors. Our model is based on a realistic bandstructure including spin-orbit coupling with electronic Hubbard-Hund interactions included at the RPA level. Spin-orbit coupling is taken into account both in the generation of spin-fluctuation mediated pairing, as well as the numerical computation of the spin susceptibility in the superconducting state. We find that spin-orbit coupling and superconductivity in conjunction can reproduce the salient experimentally observed features of the magnetic anisotropy of the neutron resonance. This includes the possibility of a double resonance, the tendency for a $c$-axis polarized resonance, and the existence of enhanced magnetic anisotropy upon entering the superconducting phase. 

\end{abstract}

\maketitle

%-------------------------------------------------------------------------------------
\section{Introduction}
\label{sec:intro}
%-------------------------------------------------------------------------------------

For unconventional superconductors, i.e., superconductors supported by non-phonon mediated Cooper pairing, the role of magnetic fluctuations has been extensively discussed in the literature~\cite{Taillefer2019,scalapinoreview}. From an empirical perspective this is motivated by the fact that the superconducting phase most often exists in close proximity to a magnetically ordered state, and significant magnetic fluctuations remain in, and potentially even generete, superconductivity. From a theoretical perspective, spin-fluctuation mediated pairing has been thoroughly studied and applied to various candidate systems including cuprates, iron-based superconductors, Sr$_2$RuO$_4$, organic Bechgaard salts, and heavy fermion materials~\cite{Taillefer2019,scalapinoreview}. This line of research goes back to the seminal theoretical work by Berk and Schrieffer in 1966~\cite{berk66}, and its extensions found in Refs.~\onlinecite{scalapino86,beal-monod86,miyake86}.

An important experimental fingerprint of the coupling between magnetic excitations and superconductivity is given by the so-called neutron spin resonance~\cite{eschrig06}. Originally discovered in the hole-doped cuprate material YBa$_2$Cu$_3$O$_{6+x}$~\cite{rossat-mignod}, the neutron resonance manifests itself as an enhanced scattering cross section in the superconducting phase at a material-specific momentum and at an energy transfer of the order of the superconducting gap scale. While the origin of the neutron resonance has been intensely discussed,~\cite{eschrig06} its most natural explanation is given in terms of an interaction-driven collective spin-1 excitation allowed by the gapped particle-hole excitations of the superconductor. Importantly, within this scenario, a sign-changing order parameter is necessary to expose the resonance. Therefore, the existence of a neutron resonance is often taken as evidence for unconventional pairing. We stress that within this picture, the neutron resonance is not evidence for pairing caused by magnetic fluctuations {\it per se}, but rather a feed-back effect of  superconductivity on the magnetic excitations.

For iron-based superconductors (FeSCs), the neutron resonance was detected early on in K-doped BaFe$_2$As$_2$~\cite{christenson2008}. Subsequently the resonance and its associated spin-gap were also found in the superconducting state of other FeSCs, and established to reside at the antiferromagnetic wavevector ${\bf Q}=(\pi,0)$. For a detailed overview of the neutron resonance studies of FeSCs we refer to the review articles in Refs.~\onlinecite{lumsden,dai,Inosov2016}. As one of the hallmarks of sign-changing gap functions, the existence of the neutron resonance was considered strong evidence for $s_{+-}$-wave pairing with sign-reversed superconducting gaps on Fermi pockets connected by ${\bf Q}=(\pi,0)$~\cite{Korshunov2008,Maier2008,hirschfeld,Korshunov2016,Korshunov2018}. This interpretation, however, was challenged, and other scenarios for the emergence of the neutron resonance feature were proposed~\cite{onari2010,takeuchi2018}. This motivated many additional studies into the properties of the neutron spin resonance including the detailed spin anisotropy of the neutron scattering resonance~\cite{dai,Inosov2016}.

Even in a non-magnetic phase significant spin anisotropy of the magnetic fluctuations exists, and can be ascribed to sizable spin-orbit coupling (SOC) in the iron-based superconducting materials~\cite{scherer2018,borisenko}. In the superconducting state, the low-energy magnetic excitations, including the neutron resonance, tend to be $c$-axis polarized as determined by spin-flip neutron scattering measurements~\cite{dai}. This is reported for a large series of compounds including LiFeAs, Ba$_{1-x}$K$_x$Fe$_2$As$_2$, BaFe$_2$(As$_{1-x}$P$_x$)$_2$, BaFe$_{2-x}$Co$_x$As$_2$, BaFe$_{2-x}$Ni$_x$As$_2$, Fe(Se,Te), FeSe, and  Sr$_{1-x}$Na$_x$Fe$_2$As$_2$.~\cite{lipscombe,Babkevich,Qureshi_LiFeAs,steffens,luo,zhang,qureshi2014,song16,song2017,ma,CZhang14,wasser,hu17,YuanLi2019} The general property of leading $c$-axis polarized low-energy susceptibility is opposite to e.g. undoped BaFe$_2$As$_2$ in the normal phase where the $a$-axis polarization dominates, in agreement with magnetic moments being aligned along the $a$ axis in the spin-density wave (SDW) phase at low temperatures. The doping-induced crossover to a dominant $c$-axis oriented susceptibility is also in agreement with out-of-plane oriented moments observed in the $C_4$-symmetric double-Q phase of Na-doped BaFe$_2$As$_2$~\cite{wasser15,christensen15}. Thus, broadly summarized, doping appears to catalyze a transition from in-plane to out-of-plane dominated low-energy magnetic fluctuations in the normal state, which are further enhanced in the superconducting state. The spin anisotropy tends to vanish in the overdoped regime.\cite{lipscombe,liu} 

Before proceeding we point out two additional aspects of the neutron resonance of FeSCs: 1) The possibility of a double resonance, and 2) the enhancement of spin anisotropy upon entering the superconducting state. The double resonance is seen, for example, in Co-doped BaFe$_2$As$_2$ and in Co-doped NaFeAs, as two separate neutron resonance peaks at low energies~\cite{steffens,Zhang_double}. The magnetic anisotropy of the double resonance was found to be much more pronounced for the lowest energy peak, which was strongly $c$-axis dominated, as compared to the higher energy peak, which was nearly isotropic in spin space~\cite{steffens,CZhang14}. Regarding the second point above, it was found, for example, in optimally doped BaFe$_2$(As$_{1-x}$P$_x$)$_2$ that the normal state appears isotropic in spin space whereas upon entering the superconducting state significant spin-space anisotropy was induced~\cite{hu17}. This is similar to BaFe$_{2-x}$Ni$_x$As$_2$ where the spin anisotropy was also observed to become further enhanced upon entering the superconducting state.~\cite{luo}.

The preferred fluctuation direction of the low-energy spin excitations is of great interest because of the potential importance of these fluctuations in determining the superconducting state. For an explicit theoretical demonstration where SOC and the associated spin anisotropy is strong enough to push the leading superconducting instability from a spin-singlet state into a spin-triplet phase, we refer to a recent theoretical study relevant to Sr$_2$RuO$_4$~\cite{astrid19}. For the case of FeSCs, a recent experimental study of Sr$_{1-x}$Na$_x$Fe$_2$As$_2$ proposed that the preferred out-of-plane $c$-axis polarized low-energy fluctuations may also be important for the competition between superconductivity and ordered in-plane magnetic SDW phases~\cite{YuanLi2019}. Thus, these works highlight the importance of understanding the polarization of the magnetic fluctuations and their role in determining the properties of the superconducting state. We note that understanding the effects of SOC in FeSCs has been also recently emphasized in terms of possible topological phases existing in these materials~\cite{hongding1,hongding2}. 

From a theoretical perspective any detailed understanding of the behavior of the magnetic anisotropy of the neutron resonance is lacking at present. The rather complex behavior of the observed anisotropy has been argued to be evidence for the importance of orbital ordering tendencies of FeSCs~\cite{luo,li}. In addition, since the ordered magnetic state exhibits $c$-axis dominated low-energy spin fluctuations, it was been suggested that the role of  antiferromagnetic SDW order may be important~\cite{Lv}. The presence of SDW order could also explain the existence of a double resonance~\cite{Lv}. The existence of the double resonance and the salient features of the magnetic anisotropy of the resonance do not, however, seem to be tied to the existence of static magnetic order. Finally we note a theoretical study of the magnetic neutron resonance comparing transverse and longitudinal fluctuations between $s_{++}$ and $s_{+-}$ superconductivity, within a simplified three-band model including only part of the SOC in the bandstructure~\cite{Korshunov_JSNM}. 

Here, we perform a theoretical study of the magnetic excitations in the superconducting phase from an itinerant weak-coupling RPA perspective, i.e., we apply a realistic ten-band model including SOC relevant to iron-based materials, and electron interactions incorporated via the multi-orbital RPA framework. In our previous paper~\cite{scherer2018}, we focussed on the paramagnetic phase and found overall agreement between the theoretical results and the experimental data in terms of material-variability, doping-, temperature-, and energy-dependence of the polarization of the low-energy magnetic fluctuations. This remarkable variability of the spin anisotropy, despite a constant atomic SOC, is a natural consequence of itinerant systems close to nesting conditions as explained in detail in Ref.~\onlinecite{scherer2018}. Here, extending this procedure to the superconducting state, we find that the main experimental findings of the magnetic anisotropy summarized above are naturally explained within our theoretical framework. In particular, the emergence of a double resonance, the tendency for a $c$-axis dominated neutron resonance, and the possibility of enhanced spin anisotropy in the superconducting phase all follow from properly including SOC in the bandstructure and the superconducting pairing. We stress that SOC is included both in the bandstructure and in the pairing kernel generated from spin-fluctuation mediated superconductivity.

The paper is structured as follows. In Sec.~\ref{sec:model}, we introduce the normal-state Hamiltonian for FeSCs and discuss details pertaining to the inclusion of SOC. While Sec.~\ref{sec:neutron} only briefly describes the approximations we employed in solving the quantum many-body problem for spin-fluctuation induced pairing in the presence of SOC and
determination of the neutron scattering amplitude of a spin-orbit coupled
superconductor, elaborations on these topics can be found in Ref.~\onlinecite{scherer2019} and the supplementary material, Sec.~\ref{sec:BdG}-~\ref{sec:RPA}, respectively. We then move to the presentation of our numerical results in Sec.~\ref{sec:results} and conclude with a discussion and possible future directions in Sec.~\ref{sec:discussion}.

%-------------------------------------------------------------------------------------
\section{Model}
\label{sec:model}
%-------------------------------------------------------------------------------------

In the following, we briefly define the Hamiltonian describing the normal metallic 
state of the spin-orbit coupled electronic system. For details regarding the Bogoliubov-de-Gennes (BdG) Hamiltonian describing the superconducting system, we refer the reader to Sec.~\ref{sec:BdG}.

We model the electronic degrees of freedom of the $3d$ shell of iron relevant for the low-energy properties of the FeSC materials by a multiorbital Hubbard Hamiltonian $ H = H_{0} - \mu_{0} N + H_{\mathrm{SOC}}  + H_{\mathrm{int}} $. Here, $H_{0}$ is the hopping Hamiltonian encoding both the electronic bandstructure in the absence of SOC and the orbital character of single-particle states and the electronic filling is fixed by the chemical potential $\mu_{0}$, with $N$ denoting the total particle number operator. Defining the fermionic operators $ c_{l i \mu \sigma}^{\dagger}$, $c_{l i \mu \sigma}$ to create and destroy, respectively, an electron on sublattice $ l $ at site $ i $ in orbital $ \mu $ with spin polarization $\sigma$, $H_{0}$ can be written as
\be
\label{eq:hopping}
H_{0} = \sum_{\sigma}\sum_{l,l^{\prime},i,j}\sum_{\mu,\nu} c_{li \mu \sigma}^{\dagger} t_{li;l^{\prime}j}^{\mu\nu} c_{l^{\prime}j \nu \sigma},
\ee
where hopping matrix elements $t_{li;l^{\prime}j}^{\mu\nu}$ are material specific. The indices $l,l^{\prime} \in \{A,B\}$ denote the 2-Fe sublattices, corresponding to the two inequivalent Fe-sites in the 2-Fe unit cell due to the pnictogen(Pn)/chalcogen(Ch) staggering about the FePn/FeCh plane, and the indices $i,j$ run over the unit cells of the square lattice. On sublattice $ l = A $, the indices $\mu,\nu$ specifying the $3d$-Fe orbitals run over the set $ \{d_{xz}, d_{yz}, d_{x^2-y^2}, d_{xy}, d_{3z^{2}-r^{2}}\}$, while on sublattice $ l = B $, we pick the gauge $ \{- d_{xz}, - d_{yz}, d_{x^2-y^2}, d_{xy}, d_{3z^{2}-r^{2}}\}$. We note, that while without SOC, a 1-Fe description (with one iron site and correspondingly 5-orbitals per unit cell) is possible, through the introduction of SOC the 2-Fe and 1-Fe descriptions are no longer unitarily equivalent. In the above-defined phase-staggered basis, the site-local SOC-Hamiltonian becomes
\be 
\label{eq:SOC}
H_{\mathrm{SOC}} = \frac{\lambda}{2}\sum_{l,i} \sum_{\mu,\nu} \sum_{\sigma,\sigma^{\prime}}
c_{l i \mu \sigma}^{\dagger} [{\bf L}_{l}]_{\mu\nu}\cdot{\boldsymbol \sigma}_{\sigma\sigma^{\prime}}c_{l i \nu \sigma^{\prime}},
\ee
with $ \boldsymbol \sigma $ the vector of Pauli matrices acting in spin space.
The components of the angular momentum operator $ [{\bf L}_{l}]_{\mu\nu} $ satisfy $ [{L}_{A}^{x,y}]_{\mu\nu} = - [{L}_{B}^{x,y}]_{\mu\nu} $, $ [{L}_{A}^{z}]_{\mu\nu} = [{L}_{B}^{z}]_{\mu\nu} $. The reason for the breakdown of the unitary equivalence of 1-Fe and 2-Fe descriptions can be found in the properties of the angular momentum operator. The unitary transformation, which in the absence of SOC block-diagonalizes \Eqref{eq:hopping} (with two blocks, where each gives rise to a 5-orbital Hamiltonian in distinct regions of momentum space), does not produce a block-diagonal angular momentum operator, i.e., \Eqref{eq:SOC} is not block-diagonalized by the same transformation. Thus, the proper inclusion of SOC necessitates either the use of a 2-sublattice representation (2-Fe description) or the introduction of a (momentum-space) non-local angular momentum operator (1-Fe description).
\begin{figure}
\centering
%-----------------------------------------------------------
\begin{minipage}{1\columnwidth}
\centering
\includegraphics[width=1\columnwidth]{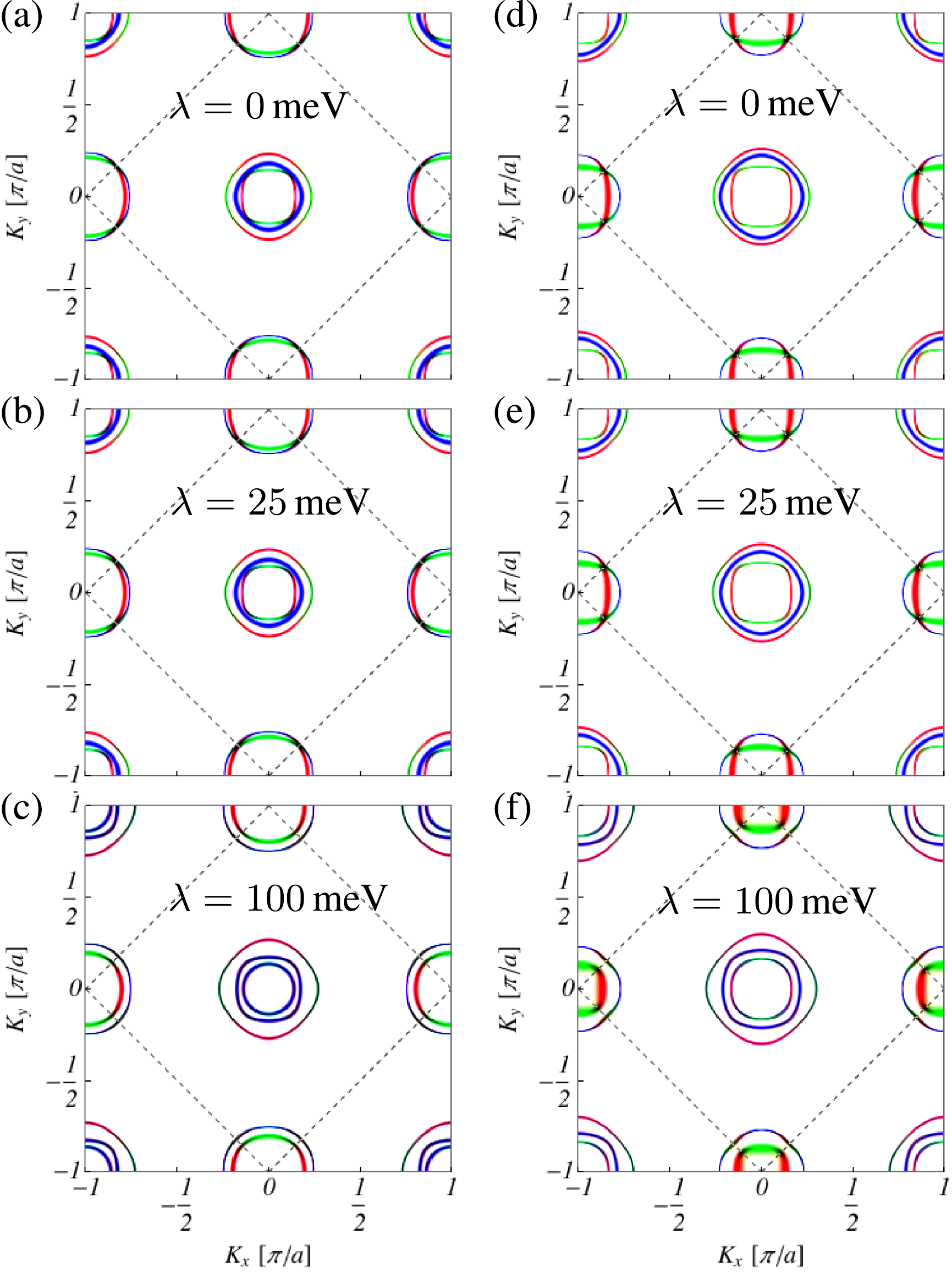}
\end{minipage}
%-----------------------------------------------------------
\caption{Normal state Fermi surfaces in the 1-Fe BZ ($(K_{x},K_{y})$ denotes momenta in the 1-Fe BZ coordinate system) extracted from the orbitally resolved contributions to the electronic spectral function with (a)-(c) $ \mu_{0} = 0 $\,eV and (d)-(e) $ \mu = -45 $\,meV for increasing $\lambda$. The 2-Fe BZ is indicated by the dashed square. The colors refer to $d_{xz}$ (red), $d_{yz}$ (green) and $d_{xy}$ (blue) orbital contributions. SOC leads to a splitting of the states at the 2-Fe BZ boundary.}
\label{fig:FS}
\end{figure}
In this work, we will use hopping parameters $t_{li;l^{\prime}j}^{\mu\nu}$ as specified in Ref.~\onlinecite{ikeda2010}. The effect of SOC on the Fermi surface and the orbital composition of Fermi surface states for two different chemical potentials is shown in Fig.~\ref{fig:FS}, where we model doping by a rigid band shift. Without SOC, the Fermi surfaces feature three hole pockets around the $\Gamma$ point and two electron pockets around the $M$ point. The inner and outer hole pockets are mostly composed of $d_{xz}$ and $d_{yz}$ orbitals, while the middle hole pocket is dominated by the $d_{xy}$ orbital. The approximate nesting of hole and electron pockets with nesting vectors ${\bf Q}_{1}$ and $ {\bf Q}_{2} $ leads to strong spin fluctuations at these wavevectors, which in an itinerant weak coupling picture eventually gives rise to an SDW instability and the condensation of SDW order. In terms of spin-fluctuation mediated superconductivity, these spin fluctuations will naturally give rise to a so-called $s_{+-}$ pairing state, where, due to the spin-mediated, repulsive interpocket interaction, the gap features a $\pi$-phase between hole and electron pockets. Typically, the superconducting instability emerges upon either hole or electron doping, when the SDW ordering tendencies are sufficiently suppressed. For strongly doped systems, the Fermi surface topology will eventually change due to the vanishing of electron or hole pockets~\cite{hirschfeld}. In these more extreme cases, different pairing states than the $s_{+-}$  can be realized due to the concomitant changes in the momentum structure of the spin-fluctuation mediated interaction. Returning to the influence of SOC on the electronic states, it is obvious from Figs.~\ref{fig:FS}(b),(c) and~\ref{fig:FS}(e),(f) that it leads to a splitting of states at the 2-Fe BZ boundary. Correspondingly, it tends to mix and equalize the orbital character of the electron pockets. The same effect seems to occur for the hole pockets.

We note here that the superconducting gap scale and the SOC-strength $\lambda$ are treated as free parameters in our model, and have been varied in a quite generous parameter interval in order to obtain a  complete picture of the possible SOC-induced spectral features of the neutron resonance mode of the superconducting system as emerging from the sign-changing $s_{+-}$ state. While the use of $\lambda $ as large as $100\,$meV might strictly speaking not be realistic, it is worth emphasizing that the bandstructure entering our calculations has a bandwidth of about $5\,$eV. While for 1111 FeSCs the bandwidth renormalization of DFT-LDA bands due to correlation effects is rather weak, these renormalizations can reduce the bandwidth by a factor 2-3 for 122 FeSCs. As it presently seems unclear, how the effective SOC energy scale is affected by inclusion of correlations on top of DFT bandstructures, we deem it a sensible strategy to explore a wide window of parameter values in order to clearly expose the role of the SOC on the neutron spin resonance. 

Completing the discussion of the model Hamiltonian, we finally turn to the interactions of the $3d$ states, which are modeled by a local Hubbard-Hund interaction term
\be
\label{eq:interaction}
H_{\mathrm{int}} & = &  
U \sum_{l,i,\mu} n_{l i \mu \uparrow} n_{l i \mu \downarrow} + 
\left(U^{\prime} \!-\! \frac{J}{2}\right) \sum_{l,i,\mu < \nu, \sigma,\sigma^{\prime}} 
\!\!\!\!\!\! n_{li \mu \sigma} n_{li \nu \sigma^{\prime}} \\
& & \hspace{-3.0em} 
- 2 J \!\! \sum_{l,i, \mu < \nu} \!\!\! {\bf S}_{li\mu}\cdot{\bf S}_{li\nu}  \! + \! 
\frac{J^{\prime}}{2}\!\!\!\! \sum_{l,i, \mu \neq \nu,\sigma} \!\!\! \left(c_{li\mu\sigma}^{\dagger}c_{li\mu\bar{\sigma}}^{\dagger}c_{li\nu\bar{\sigma}}c_{li\nu\sigma}+\mathrm{h.c.}\right). \nn
\ee
The Hamiltonian \Eqref{eq:interaction} is parametrized by an intraorbital Hubbard $U$, an interorbital coupling $U^{\prime}$, Hund's coupling $J$ and pair hopping $J^{\prime}$, satisfying $U^{\prime} = U - 2J$, $J = J^{\prime}$ due to orbital rotational invariance of the Coulomb matrix elements with respect to the Wannier basis functions. The operators for local charge and spin are $n_{li\mu} = n_{li\mu\uparrow} + n_{li\mu\downarrow}$ with $n_{li\mu\sigma} = c_{li\mu\sigma}^{\dagger} c_{li\mu\sigma}$ and ${\bf S}_{li\mu} = 1/2\sum_{\sigma\sigma^{\prime}} c_{li\mu\sigma}^{\dagger} {\boldsymbol \sigma}_{\sigma\sigma^{\prime}}c_{li\mu\sigma^{\prime}}$, respectively. In the following, we will further constrain the value of the Hund's coupling to $ J = U/4 $ in order to reduce the number of parameters. Below, we will treat interaction effects at the level of the RPA. The bare interaction vertex defined by the interaction Hamiltonian above will provide the corresponding approximation to the 2-particle irreducible (2PI) vertex in the particle-hole channel.

%-------------------------------------------------------------------------------------
\section{Neutron Scattering Amplitude}
\label{sec:neutron}
%-------------------------------------------------------------------------------------

To model the superconducting properties of the system, we proceed as follows. First, in order to take into account SOC already at the level of the spin-fluctuation mediated pairing mechanism, we construct the 2PI vertex in the particle-particle channel by performing the RPA resummation of particle-hole diagrams, employing the bare 2PI particle-hole vertex and normal-state Greens functions including the effect of SOC. We then determine the leading solution of the corresponding Fermi-surface projected  linearized gap equation (LGE) (Bethe-Salpeter equation in the particle-particle channel) in the presence of SOC (for details, see Ref.~\onlinecite{scherer2019}) for given $\lambda$ and interaction parameters $U$, $J$. We note that we solve the LGE in the static approximation, i.e., we take the 2PI particle-particle vertex at vanishing energy arguments. 

Due to the locking of spin and orbital degrees of freedom by virtue of the SOC Hamiltonian, the pairing problem for Fermi surface states is most naturally formulated for Cooper pairs composed of the states in a Kramer's doublet, i.e., single-particle states at momenta ${\bf k}$ and $-{\bf k}$ which are related by time-reversal. In fact, due to time-reversal and inversion symmetry in the normal state, each band is still doubly degenerate. Due to breaking of continuous spin-rotational symmetry by SOC, spin no longer represents a good quantum number to label the states in this degenerate subspace. It is possible, however, to define a pseudo-spin degree of freedom, which coincides with physical spin as $ \lambda \to 0 $, and which indeed has the transformation properties of a spin-1/2 degree of freedom, even for finite SOC. The solutions of the Fermi surface projected LGE can therefore be classified as even-parity pseudo-spin singlet and odd-parity pseudo-spin triplet solutions. As mentioned in Sec.~\ref{sec:model}, the strong spin fluctuations at the nesting vectors tend to drive a Cooper instability to a superconducting state with even-parity $ s_{+-} $ gap structure upon doping. In this work, we will therefore exclusively be concerned with the neutron-scattering signatures of $s_{+-}$ gap solutions. By restricting the doping range such that the topology of the Fermi surface does not drastically change, the spin fluctuations at wavevectors $ {\bf Q}_{1} $ and $ {\bf Q}_{2} $ are indeed dominant and entail a leading Cooper instability of even-parity $s_{+-}$-type.

The LGE solutions obtained for fixed $ \lambda $ and interaction parameter $U$ then serve as the starting point for defining a BdG Hamiltonian (see Sec.~\ref{sec:BdG}) and the corresponding Nambu-Gorkov Greens function (see. Sec.~\ref{sec:NG}) for the superconducting state. Here, we make use of the procedure described in Ref.~\onlinecite{maier2009} in order to obtain gap solutions throughout the entire Brillouin zone from the Fermi-surface projected LGE. We denote the corresponding pseudo-spin singlet gap function by
\be 
\hat{\Delta}_{0}^{b}({\bf k}) = \Delta_{0} \, g^{b}({\bf k}),
\ee 
with $ \Delta_{0} $ a parameter fixing the gap amplitude, and $ g^{b}({\bf k}) $ dimensionless functions (one for each band $b$) defined on the 2-Fe BZ, describing the gap structure obtained from the LGE. The functions $ g^{b}({\bf k}) $ are normalized, such that $ \Delta_{0} $ is the maximum value of $ |  \hat{\Delta}_{0}^{b}({\bf k})  | $ (where maximization is carried out over  ${\bf k}$ and $b$). We note that we do not model a temperature-dependent $ \Delta_{0} $, but instead use $ \Delta_{0} $ as a free parameter.

Again approximating the 2PI particle-hole vertex by the bare vertex defined by \Eqref{eq:interaction}, we finally determine the RPA spin susceptibility of the superconducting state, see Sec.~\ref{sec:RPA} for details. In the following, we will refer to this approximation as BCS+RPA~\cite{bulut1992,bulut1991}. To this end, we compute the connected, imaginary-time spin-spin correlation function in the superconducting state (here $i,j$ refer to the spatial directions $x ,y, z$)
\be
\label{eq:susctensor}
\!\!\!\!\!\!\! \chi^{ij}(\mathrm{i}\omega_n,{\bf q}) \! = \! g^{2} \!\! \int_{0}^{\beta} \!\! d\tau \,
\mathrm{e}^{\mathrm{i}\omega_n \tau}
\langle \mathcal{T}_{\tau} S^{i}_{{\bf q}}(\tau) S^{j}_{-{\bf q}}(0)\rangle_{\scriptsize{c,\mathrm{BCS}+\mathrm{RPA}}},
\ee
with $g=2$ and the Fourier-transformed electron spin operator (in the imaginary-time Heisenberg picture) for the 2-Fe unit cell given as
\be
S^{i}_{{\bf q}}(\tau) = \frac{1}{\sqrt{\mathcal{N}}}\sum_{{\bf k},l,\mu,\sigma,\sigma^{\prime}} c_{{\bf k} - {\bf q}l\mu\sigma}^{\dagger}(\tau) \frac{\sigma_{\sigma\sigma^{\prime}}^{i}}{2}c_{{\bf k}l\mu\sigma^{\prime}}(\tau).
\ee
The symbol $\mathcal{T}_{\tau}$ in \Eqref{eq:susctensor} denotes the time-ordering operator for the imaginary time variable $\tau$ and $ c_{{\bf k}l\mu\sigma}^{\dagger} $, $ c_{{\bf k}l\mu\sigma}$ denote the Fourier-transformed fermionic creation and annihilation operators, respectively, and $ \mathrm{i}\omega_n$ denotes a bosonic Matsubara frequency. The momentum ${\bf k}$ is an element of the 2-Fe BZ, and $\mathcal{N}$ counts the number of 2-Fe unit cells. Performing analytic continuation $ \mathrm{i} \omega_{n} \to \omega + \mathrm{i} \eta $ of $ \chi^{ij}(\mathrm{i}\omega_n,{\bf q}) $ in \Eqref{eq:susctensor} with $ \eta > 0 $, we gain access to the momentum- and frequency-resolved spectral density of magnetic excitations with different spin-space polarization as probed by polarized neutron scattering. Due to the proportionality of the experimentally accessible neutron scattering amplitude and the spectral density of magnetic excitations, the theoretical determination of the magnetic susceptibility in the superconducting state is a valuable tool in unveiling the underlying electronic structure.

\begin{figure*}[t!]
\centering
%-----------------------------------------------------------
\begin{minipage}{1\textwidth}
\centering
\includegraphics[width=1\textwidth]{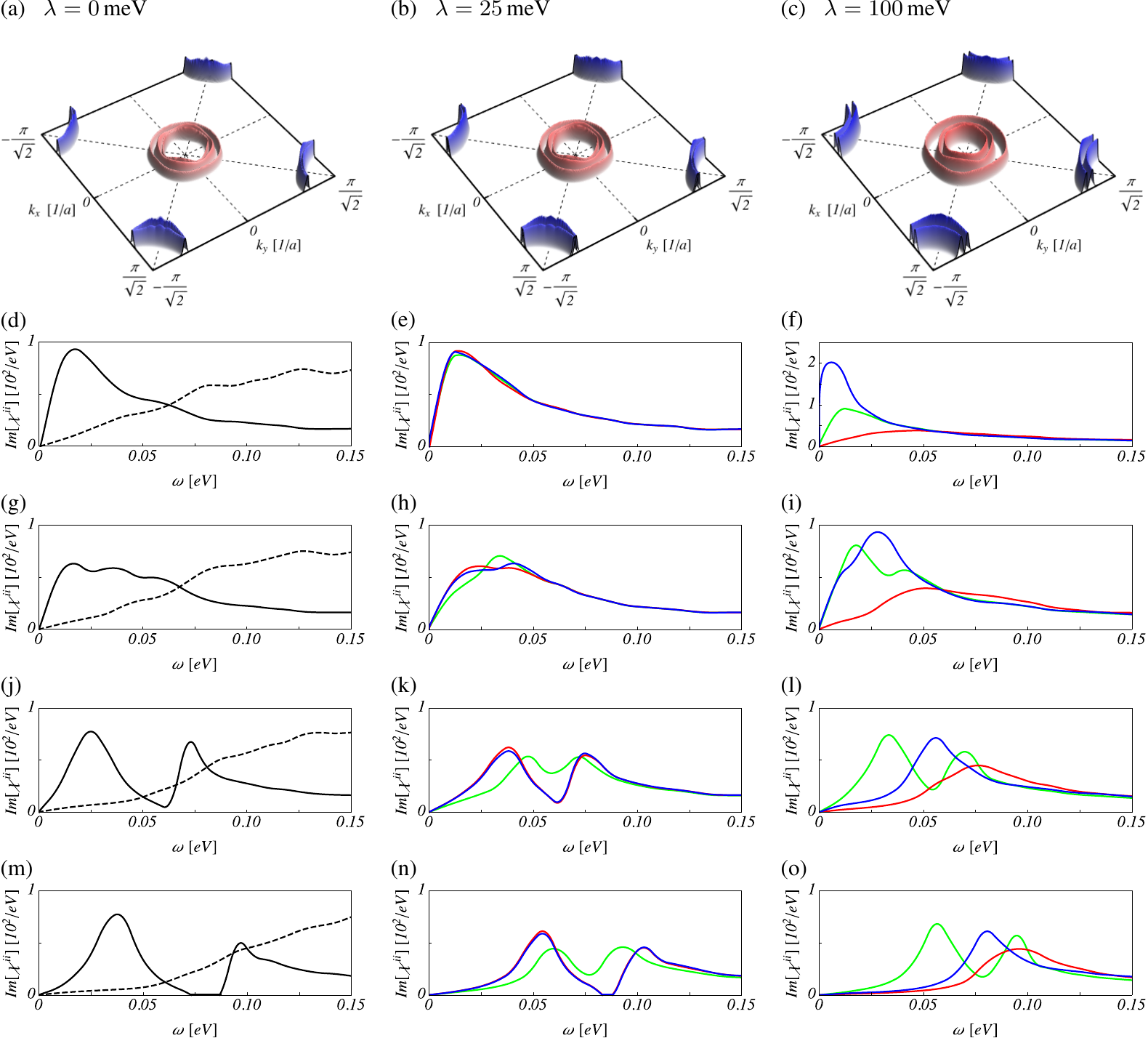}
\end{minipage}
%-----------------------------------------------------------
\caption{(a)-(c) (Pseudo-)Spin singlet $s_{+-}$ gap function $ \hat{\Delta}_{0}^{b}({\bf k}) $ in the 2-Fe BZ as obtained from the linearized gap equation with a 2PI pairing vertex in RPA approximation for various values of spin-orbit coupling strength $\lambda$ at chemical potential $\mu_{0} = 0 \, $meV. Red and blue color correspond to positive and negative gap values, respectively. (d)-(o) Imaginary part of the susceptibility at the nesting vector ${\bf Q}_{1}$ as a function of energy for varying $\lambda$ and gap amplitude $\Delta_{0}$. The SOC-strength $\lambda$ is fixed in each column and corresponds to the values in (a)-(c). 
The gap amplitude $\Delta_{0}$ varies as (d)-(f) $\Delta_{0} = 0\,$meV, (g)-(i) $\Delta_{0} = 25\,$meV, (j)-(l) $\Delta_{0} = 50\,$meV and (m)-(o) $\Delta_{0} = 75\,$meV.
Blue color corresponds to polarization $ i = x $, red to $ i = y $ and green to $ i = z $. 
The black dashed curves in (d), (g), (j), (m) show the non-interacting susceptibilities (scaled up for visibility by a factor of 75).}
\label{fig:res1}
\end{figure*}

%-------------------------------------------------------------------------------------
\section{Results}
\label{sec:results}
%-------------------------------------------------------------------------------------

%
In the following, we will present the results for the neutron scattering amplitude
as extracted from the imaginary part of the spin susceptibility determined in BCS+RPA.
As we are interested in the spectral splitting and magnetic anisotropy of the neutron
resonance mode of the spin fluctuation induced $s_{+-}$ state in the presence of SOC, 
we focus on the frequency dependence of the neutron scattering amplitude at 
wavevector $ {\bf Q}_{1} = (\pi,0) $. Here we choose a coordinate system with $ x = a $, $ y = b $ and $ z = c $, where the cartesian axes are aligned with the orthorhombic crystal axes. We further note that in tetragonal states, which we consider here, the magnetic response at the wavevector $ {\bf Q}_{2} = (0,\pi) $ is related to the response at $ {\bf Q}_{1} $ by a $C_{4}$ rotation in the $ab$ plane. The cross-terms with $ i \neq j $ in Eq.~\ref{eq:susctensor} vanish for tetragonal systems.

We first discuss the results obtained for a chemical potential $\mu_{0} = 0 \,$eV, for which the system has an electronic filling of $ n = 6 $ (with $ n $ the number of filled states per unit cell).
We obtained the LGE solutions for a range of $ U $ and $ \lambda $ values. The leading instability in the Cooper channel was, as expected from the momentum structure of the susceptibility, always found to be of even parity $s_{+-}$ type. The resulting gap structures $g^{b}({\bf k})$ for $ U = 0.80\,$eV and various values for $\lambda$ are shown in Fig.~\ref{fig:FS}(a)-(c). The value of the interaction was chosen to bring the system close to the SDW instability. Evaluating the susceptibility for the BCS Hamiltonian with the corresponding (pseudo-)spin singlet, intraband order parameter $ \hat{\Delta}_{0}^{b}({\bf k}) $, we clearly observe the formation of neutron resonance features in the spectral density of magnetic excitations upon going from the normal to the superconducting state, see Fig.~\ref{fig:res1}. We note that we took the liberty to tune the interaction parameter in the superconducting state to a slightly larger value of $ U = 0.88 \, $eV in order to enhance the resonance signatures. Qualitatively, however, the results remain unchanged compared to using the interaction value used in the construction of the pairing vertex entering the LGE. 
\begin{figure*}[t!]
\centering
%-----------------------------------------------------------
\begin{minipage}{1\textwidth}
\centering
\includegraphics[width=1\textwidth]{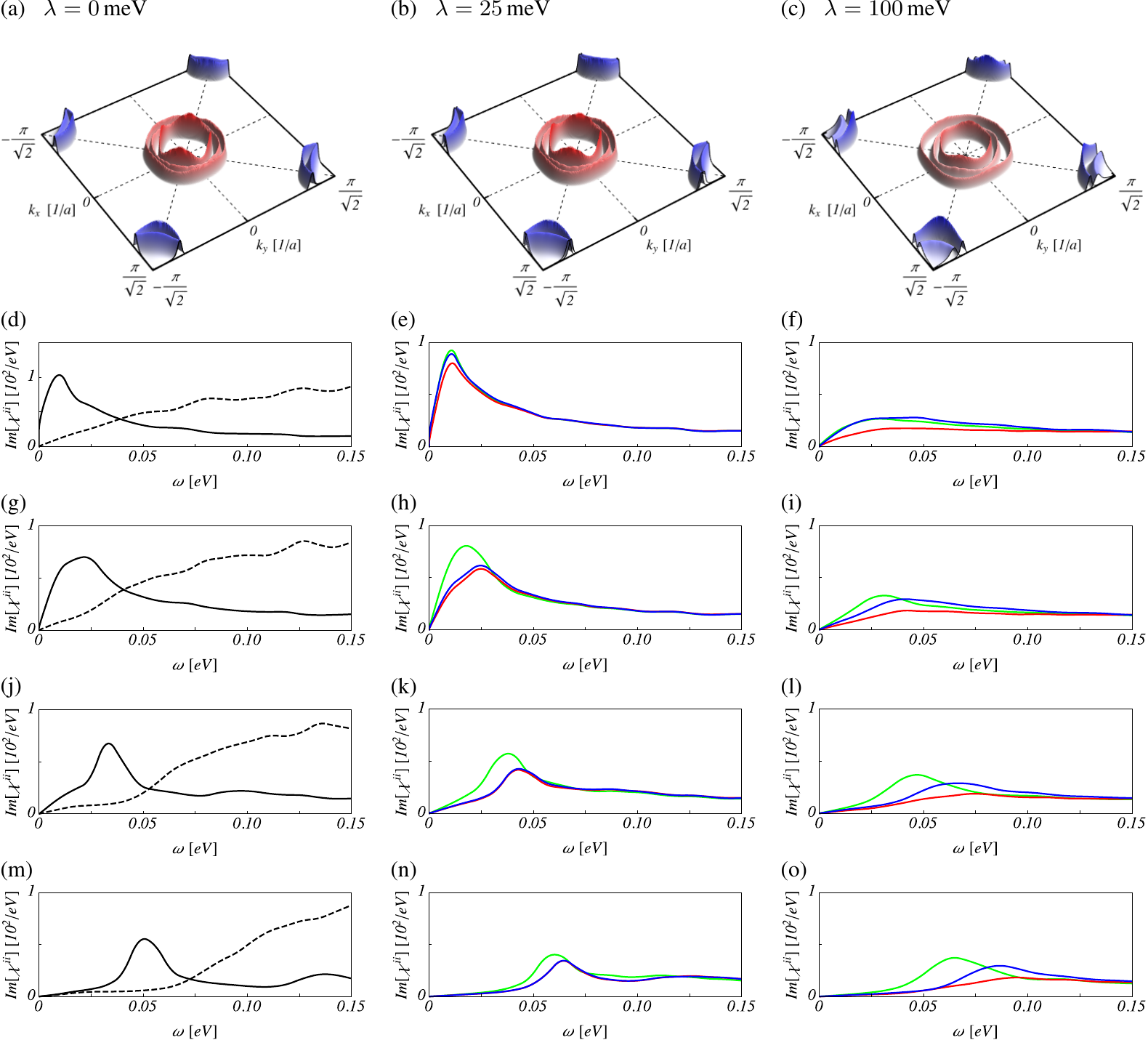}
\end{minipage}
%-----------------------------------------------------------
\caption{(a)-(c) (Pseudo-)Spin singlet $s_{+-}$ gap function $ \hat{\Delta}_{0}^{b}({\bf k}) $ in the 2-Fe BZ as obtained from the linearized gap equation with a 2PI pairing vertex in RPA approximation for various values of spin-orbit coupling strength $\lambda$ at chemical potential $\mu_{0} = - 45 \, $meV, corresponding to a hole-doped system. Red and blue color correspond to positive and negative gap values, respectively. (d)-(o) Imaginary part of the susceptibility at the nesting vector ${\bf Q}_{1}$ as a function of energy for varying $\lambda$ and gap amplitude $\Delta_{0}$. The SOC-strength $\lambda$ is fixed in each column and corresponds to the values in (a)-(c). The gap amplitude $\Delta_{0}$ varies as (d)-(f) $\Delta_{0} = 0\,$meV, (g)-(i) $\Delta_{0} = 25\,$meV, (j)-(l) $\Delta_{0} = 50\,$meV and (m)-(o) $\Delta_{0} = 75\,$meV.
Blue color corresponds to polarization $ i = x $, red to $ i = y $ and green to $ i = z $. The black dashed curves in (d), (g), (j), (m) show the non-interacting susceptibilities  (scaled up for visibility by a factor of 75).}
\label{fig:res2}
\end{figure*}
In Fig.~\ref{fig:res1}(d)-(o) we vary $\Delta_{0}$ from $0$ (d)-(f) over $25$ (g)-(i) and $50$ (j)-(l) to $75\,$meV (m)-(o). The low-energy neutron scattering amplitude in the superconducting state is decreased compared to the normal state. Due to the sign-changing nature of the $s_{+-}$ state, however, the emergence of a bound state in the form of the neutron resonance mode is possible for energies $ \omega < 2 \Delta_{0} $. 

Momentarily focussing on the case without SOC ($\lambda=0$), the neutron scattering amplitude is isotropic, i.e., as expected it exhibits no difference between the different $x$, $y$ and $z$ polarization channels, see Fig.~\ref{fig:res1}(d),(g),(j),(m). We further observe an apparent splitting of the resonance mode upon increasing $ \Delta_{0} $, while the positions of the resonance peaks simultaneously shift to higher energies. 

Increasing SOC in the normal state, the polarization resolved scattering amplitudes eventually split, Fig.~\ref{fig:res1}(d)-(f). The splitting increases with SOC and Fig.~\ref{fig:res1}(f) clearly demonstrates that the paramagnetic spin excitations with $x$ polarization have become almost gapless, while paramagnon excitations with $y$ and $z$ polarization reside at larger energies with fluctuations polarized along $y$ exhibiting the highest energy paramagnon branch. Moving on to the superconducting state we observe qualitatively different behavior of the neutron scattering amplitude with energy in the small and large SOC regimes. With $ \Delta_{0} = 25\,$meV, only a small polarization dependent differentiation is visible for $ \lambda =  25$\,meV, much like in the paramagnetic state (Fig.~\ref{fig:res1}(h)). The large SOC case, however, features a two-peak structure in the $z$-polarization channel, with a third peak in the $x$-polarization channel situated between the two $z$-polarization peaks on the energy axis (Fig.~\ref{fig:res1}(i)). This feature grows more pronounced as the superconducting gap amplitude is increased, see Fig.~\ref{fig:res1}(l),(o). On the small SOC side, increasing the gap amplitude beyond $25\,$meV eventually leads to several split peaks as well. Here, the $x$- and $y$-polarized excitation modes occur with almost equal spectral weight in a double-peak structure, while the $z$-polarized mode also features a double peak structure, which is squeezed between the peaks of the other two polarization channels, see Fig.~\ref{fig:res1}(k),(n). It seems rather noteworthy that the magnetic anisotropy of the neutron resonance mode(s) is not simply inherited from the paramagnetic state, but instead seems to emerge from the interplay of the $s_{+-}$ state and SOC. 

Before we move to the results obtained for negative chemical potential, we emphasize that the case of $ \mu = 0 \, $eV is relevant for both parent and doped FeSC materials in terms of the magnetic anisotropy of the normal state. Previsouly, we have identified that dominating $x$ polarization of the normal state is a robust feature in a certain window around $ \mu = 0 \, $eV~\cite{scherer2018}.

Moving to the case of chemical potential $\mu_{0} = -0.45\,$meV, resembling a hole-doped system with dominating $z$ polarization in the normal state, we obtained LGE solutions for an interaction parameter $ U = 0.70\,$eV, which are shown in Fig.~\ref{fig:res2}(a)-(c). In contrast to the $\mu = 0 \,$eV system, hole doping leads to a stronger gap-anisotropy, in that the gap on the inner hole pocket now has a markedly larger maximal amplitude than the middle and outer hole pockets. The gap on the inner hole pocket also turns out larger than on the electron pockets. The gap on the outer electron pocket features a much larger gap variation than the inner electron pocket as the Fermi momentum paces out the pocket shape. Increasing SOC seems to have two effects: While the interpocket anisotropy eventually decreases for the hole pockets, the gap amplitude on the outer electron pocket develops a more pronounced maximum (with the gap on the inner electron pocket being basically unaffected by SOC). 

Performing the same analysis as before (where we tuned the interaction to $U = 0.80\,$eV for the same purpose as above), we observe the emergence of a resonance feature upon increasing the gap amplitude, see Fig.~\ref{fig:res2}(d),(g),(j),(m). In contrast to the $ \mu = 0\,$eV system, however, in the absence of SOC we observe only a single peak in the magnetic excitation spectrum. Switching on SOC, the normal state features dominant and almost gapless $z$-polarized paramagnon excitations. The splitting in paramagnon gaps is, however, rather small. Consequently, the energy-dependent spectral weight of the three polarization channels is almost the same, see Fig.~\ref{fig:res2}(e),(f). We note that we have chosen the chemical potential to provide the \textit{largest} possible splitting between $x$- and $z$-polarized paramagnons. Increasing the gap amplitude leads to a stronger splitting of the spectral weight contributions, even for fixed SOC. In the case of a small SOC energy scale, the $x$- and $y$-polarization channels behave almost identically, cf. Fig.~\ref{fig:res2}(h),(k),(n), while larger SOC leads to a more pronounced differentiation of these two channels, cf. Fig.~\ref{fig:res2}(f),(i),(l),(o). We note further, that in the case of large SOC, apparently no resonance forms in the $y$-polarization channel. It is only the $z$- and $x$-polarized spectral weight curves that feature more or less well formed peaks, with the peak corresponding to $z$-polarized excitations systematically occuring at lower energies.

%-------------------------------------------------------------------------------------
\section{Discussion and Conclusions}
\label{sec:discussion}
%-------------------------------------------------------------------------------------

Having established, that a weak coupling BCS+RPA calculation can 
produce a variety of realizations of neutron-resonance features, all
depending on the detailed values of chemical potential, SOC
strength and gap amplitude, we attempt to uncover hints pointing to the
mechanism behind the generation of the observed anisotropy.

Here it is worth emphasizing that the interaction Hamiltonian $ H_{\mathrm{int}} $ cannot generate anisotropy on its own, as it respects rotational symmetry in both orbital and spin space. This leaves as the only two factors influencing the anisotropy in the neutron scattering amplitude i) the interplay of SOC with the bandstructure in the superconducting system and ii) the spin structure of the superconducting solutions. As a detailed understanding of i) turned out to be a rather involved problem even in the normal state~\cite{scherer2018}, we do not attempt to provide a full answer here. Rather, we try to assess the explanatory possibilities of ii), essentially by testing for the effect of the spin-triplet component of the superconducting order parameter in orbital $ \otimes $ spin space on the magnetic anisotropy. To this end, we performed additional numerical calculations, where we either projected out the spin-triplet component in the $ \bar{\hat{\Delta}}_{0}^{b}({\bf k}) \hat{\Delta}_{0}^{b^{\prime}}({\bf k} - {\bf q}) $ prefactor of the $ \bar{F}F $ contribution to the irreducible bubble in the particle-hole channel but left the electronic structure of the superconductor unaltered, or where we removed the spin-triplet component entirely, thereby also altering the gap structure. In both cases, it turns out that the removal of the spin-triplet component does not lead to qualitatively different results than those summarized in Fig.~\ref{fig:res1} and Fig.~\ref{fig:res2}. This observation can partly be explained by noting that the spin-triplet component of the superconducting order parameter is basically driven by SOC. Since SOC (compared to the electronic bandwidth) is a rather small energy scale for the systems under consideration, the largest gap amplitudes of the spin-triplet component turn out to be suppressed by at least one order of magnitude with respect to the spin-singlet component.

As the spin-triplet component and its associated $d$-vector do apparently not play a decisive role in determining the magnetic anisotropy of the superconducting state at the nesting vectors $ {\bf Q}_{1} $ and $ {\bf Q}_{2} $, we conclude that it must be the subtle interplay of SOC and the bandstructure of the $ s_{+-} $ state that gives rise to an inherent tendency of the $ s_{+-} $ state to favor $z$-axis polarized excitations at low energies. Indeed, as a comparison of the results for $ \mu = 0 \, $meV and $ \mu = - 45\,$meV systems shows, a large SOC is required to render the low-energy magnetic excitations to be $z$ polarized, if the normal state paramagnons at low energies are preferably $x$ polarized. For the case of dominantly $z$ polarized paramagnons (with $x$, $y$ polarized excitations having larger, yet similarly sized excitation gaps) the $s_{+-}$ order parameter and increased SOC have the tendency to enhance the normal state magnetic anisotropy.
We emphasize once more, that, as the anisotropy is not generated by the interaction, the results we found are -- at least at a qualitative level -- robust with respect to variations of the interaction strength. We do, however, find pronounced changes in the relative spectral weight distribution of the different polarization channels, when the superconducting system approaches an SDW instability by tuning the interaction appropriately. Typically, the lowest-energy peak grows considerably in size, while simultaneously moving to lower energies. The remaining peaks only show a moderate increase in spectral weight and are more or less locked in place on the energy axis. 

In the introduction we emphasized a number of experimental observations that broadly summarized the main findings of the spin anisotropy of the neutron resonance. These included the possibility of a double resonance, the tendency for a $c$-axis dominated resonance, the enhancement of spin anisotropy below $T_c$, and the diminishing anisotropy for large enough doping levels. While the latter point has not been investigated numerically in this work, we do know from our earlier study that there is a clear tendency of the normal state spin anisotropy to vanish for sufficiently large doping levels~\cite{scherer2018}. We expect the same to hold true for the anisotropy of the neutron resonance. Regarding the former three points, however, they seem to be reasonably well captured by the BCS+RPA approach, including SOC, as presented in this work. It is simply an inherent property of the bandstructure of the FeSCs and the form of the SOC in these materials, that the low-energy resonance favors $c$-axis polarization. Further, it is a property of a large enough superconducting gap to secure the possible existence of a double resonance. Finally, the interplay of both SOC and superconductivity cause the enlarged spin anisotropy upon entering the superconducting phase. From this qualitative agreement between the model and the experimental situation we conclude that the methodology presented in this work is sufficient to explain the properties of the low-energy spin anisotropy of FeSCs. 

An interesting future direction includes more material-specific modelling, and to address the challenge of quantitative agreement between measurements and theory. This requires not only the necessity to go beyond BCS+RPA in terms of electronic interaction effects, but also {\it {ab initio}} developments allowing for quantitatively matching bandstructures as a stating point. Another interesting study would be to investigate superconductors with larger SOC than in FeSCs. In that case, perhaps, the spin texture of the superconducting order parameter alone could more strongly influence the magnetic anisotropy of the low-energy spin excitations in general, and the neutron spin resonance in particular.

%-------------------------------------------------------------------------------------
\begin{acknowledgments}
We acknowledge financial support from the Carlsberg Foundation. 
\end{acknowledgments}
%-------------------------------------------------------------------------------------

%-------------------------------------------------------------------------------------

%-------------------------------------------------------------------------------------

%%-------------------------------------------------------------------------------------
%% Supplementary Material
%%-------------------------------------------------------------------------------------

\begin{widetext}

\newpage

%%%%%%%%%% Merge with supplemental materials %%%%%%%%%%
%%%%%%%%%% Prefix a "S" to all equations, figures, tables and reset the counter %%%%%%%%%%
\setcounter{equation}{0}
\setcounter{section}{0}
\setcounter{figure}{0}
\setcounter{table}{0}
\setcounter{page}{1}
\makeatletter
\renewcommand{\theequation}{S\arabic{equation}}
\renewcommand{\thesection}{S\arabic{section}}
\renewcommand{\thefigure}{S\arabic{figure}}
\renewcommand{\bibnumfmt}[1]{[S#1]}
\renewcommand{\citenumfont}[1]{S#1}

\begin{center}
\textbf{\large Supplementary Material: ``Effects of spin-orbit coupling on the neutron spin resonance in iron-based superconductors''}
\end{center}

%-------------------------------------------------------------------------------------
\section{BdG Hamiltonian}
\label{sec:BdG}
%-------------------------------------------------------------------------------------

In the following, we will provide details on the Bogoliubov-de-Gennes (BdG) Hamiltonian for a spin-orbit coupled superconductor and the corresponding Nambu-Gorkov Greens function entering the computation of maginary-time correlation functions in the superconducting state. We start out with transforming the normal-state Hamiltonian to momentum space, where we define the Fourier-transformed fermionic operators
\be 
c_{{\bf k}l\mu\sigma} = \frac{1}{\sqrt{\mathcal{N}}} \sum_{i} \mathrm{e}^{\mathrm{i}{\bf k}\cdot({\bf r}_{i}+\boldsymbol{\delta}_{l})} \, c_{li\mu\sigma}, \quad c_{{\bf k}l\mu\sigma}^{\dagger} = \frac{1}{\sqrt{\mathcal{N}}} \sum_{i} \mathrm{e}^{-\mathrm{i}{\bf k}\cdot({\bf r}_{i}+\boldsymbol{\delta}_{l})} \, c_{li\mu\sigma}^{\dagger},
\ee
where the sum runs over lattice sites, ${\bf r}_{i}$ denotes the lattice vector corresponding to the $i$th unit cell and $\boldsymbol{\delta}_{l}$ denotes an intra unit cell vector referencing the relative positions of the two Fe atoms. The non-interacting Hamiltonian $ H_{0} + H_{\mathrm{SOC}} $, see \Eqref{eq:hopping} and \Eqref{eq:SOC}, can then be written as
\be 
\label{eq:hamiltonian}
H_{0} + H_{\mathrm{SOC}} = \sum_{{\bf k},l,l^{\prime},\sigma,\sigma^{\prime}} 
c_{{\bf k}l\mu\sigma}^{\dagger} \,
[h({\bf k})]_{l\mu\sigma,l^{\prime}\mu^{\prime}\sigma^{\prime}} \,
c_{{\bf k}l^{\prime}\mu^{\prime}\sigma^{\prime}},
\ee
with $ h({\bf k}) $ denoting a $20 \times 20$ ${\bf k}$-dependent Bloch-matrix acting in the single-particle Hilbert space of sublattice $\otimes$ orbital $\otimes$ spin. The Bloch-matrix $ h({\bf k}) $ is diagonalized by a unitary transformation
\be 
\label{eq:transformation}
c_{{\bf k}l\mu\sigma} = \sum_{b,\kappa} [\mathcal{U}({\bf k})]_{l\mu\sigma,b\kappa} \Phi_{{\bf k}b\kappa}, \quad
c_{{\bf k}l\mu\sigma}^{\dagger} = \sum_{b,\kappa} [\mathcal{U}({\bf k})]_{l\mu\sigma,b\kappa}^{\ast} \Phi_{{\bf k}b\kappa}^{\dagger}
\ee
with the unitary matrix $ \mathcal{U}({\bf k}) $ comprising the eigenvectors of $ h({\bf k}) $, with corresponding eigenvalues $ \epsilon_{b\kappa}({\bf k}) $. Here, the index $ b $ labels electronic bands, while $ \kappa \in \{ +, - \}$ labels a pair of degenerate states, that is guaranteed to exist at a given ${\bf k}$ for a time-reversal-invariant and inversion-symmetric system. In the solution of the pairing problem with spin-orbit coupling, the $\kappa$-degree of freedom plays the role of a pseudospin. With a judicious choice of the eigenbasis in the degenerate subspaces, the pseudospin indeed has the transformation properties of a spin-1/2 degree of freedom~\cite{SMscherer2019}.  

Moving to the superconducting state by adding a Hamiltonian describing the coupling of electrons to the pairing field, $H_{\Delta}$, the full BCS Hamiltonian containing the superconducting order parameter can be written as
\be 
\label{eq:BCS}
H_{\mathrm{BCS}} = H_{0} - \mu_{0} N + H_{\mathrm{SOC}} + H_{\Delta} = \frac{1}{2}\sum_{{\bf k}} \Psi_{{\bf k}l\mu}^{\dagger} [ h_{\mathrm{BdG}}({\bf k}) ]_{l\mu,l^{\prime}\mu^{\prime}} \Psi_{{\bf k}l^{\prime}\mu^{\prime}},
\ee
where we introduced the Nambu-Gorkov spinors $\Psi_{{\bf k}l\mu}$, $\Psi_{{\bf k}l\mu}^{\dagger}$. In terms of the original electron operators in the orbital Wannier basis, these are defined as
\be 
\Psi_{{\bf k}l\mu} = 
\left( c_{{\bf k}l\mu\uparrow}, c_{{\bf k}l\mu\downarrow}, c_{-{\bf k}l\mu\uparrow}^{\dagger}, c_{-{\bf k}l\mu\downarrow}^{\dagger} \right)^{T}, \quad
\Psi_{{\bf k}l\mu}^{\dagger} = 
\left( c_{{\bf k}l\mu\uparrow}^{\dagger}, c_{{\bf k}l\mu\downarrow}^{\dagger}, c_{-{\bf k}l\mu\uparrow}, c_{-{\bf k}l\mu\downarrow} \right),
\ee
where the transposition turns the row- into a columnvector, but acts trivially on the Fock-space operators. For each pair of indices $ l, \mu $ and $ l^{\prime }, \mu ^{\prime} $, $ [h_{\mathrm{BdG}}({\bf k}) ]_{l\mu,l^{\prime}\mu^{\prime}} $ is a $ 4 \times 4 $ matrix acting in particle-hole $\otimes$ spin-space. We can compactly write
\be 
\label{eq:BdG}
h_{\mathrm{BdG}}({\bf k})  = 
\mathcal{P}_{+} \otimes  \left( h({\bf k}) - \mu_{0} \mathbbm{1} \right)+
\mathcal{P}_{-} \otimes \left( -h^{T}(-{\bf k}) + \mu_{0} \mathbbm{1} \right)  +
\tau_{+} \otimes  \Delta({\bf k}) + \tau_{-}  \otimes  \bar{\Delta}({\bf k}),
\ee
where $ \mathcal{P}_{\pm} = \frac{1}{2} \left( \mathbbm{1} \pm \tau_{z} \right) $, 
$ \tau_{\pm} = \frac{1}{2} \left( \tau_{x} \pm \mathrm{i}\tau_{y} \right) $ with $ \tau_{i} $, $i = x,y,z$ Pauli matrices acting in particle-hole space. We keep a general orbital and spin structure for the pairing fields $ \Delta({\bf k}) $, $ \bar{\Delta}({\bf k}) $. The spin structure can be decomposed into spin-singlet ($\varsigma = 0$) and spin-triplet ($\varsigma = x, y, z$) as
\be
[\Delta({\bf k})]_{l\mu\sigma,l^{\prime}\mu^{\prime}\sigma^{\prime}} = 
\sqrt{2} \sum_{\varsigma} s_{\varsigma}  [\Delta_{\varsigma}({\bf k})]_{l\mu,l^{\prime}\mu^{\prime}} [\Gamma_{\varsigma}]_{\sigma,\sigma^{\prime}}, \quad
[\bar{\Delta}({\bf k})]_{l\mu\sigma,l^{\prime}\mu^{\prime}\sigma^{\prime}} = 
\sqrt{2} \sum_{\varsigma} [\bar{\Delta}_{\varsigma}({\bf k})]_{l\mu,l^{\prime}\mu^{\prime}} [\Gamma_{\varsigma}]_{\sigma,\sigma^{\prime}},
\ee
where we introduced the Balian-Werthamer (BH) spin matrices 
\be 
\Gamma_{0} = \frac{1}{\sqrt{2}}\mathbbm{1}\mathrm{i}\sigma_{y}, \quad
\Gamma_{x} = \frac{1}{\sqrt{2}}\sigma_{x}\mathrm{i}\sigma_{y}, \quad
\Gamma_{y} = \frac{1}{\sqrt{2}}\sigma_{y}\mathrm{i}\sigma_{y}, \quad
\Gamma_{z} = \frac{1}{\sqrt{2}}\sigma_{z}\mathrm{i}\sigma_{y},
\ee
characterizing the spin structure of singlet and triplet Cooper pairs, and $ s_{\varsigma} = +1 $ for $ \varsigma \in \{ x,z \} $ and $ s_{\varsigma} = -1 $ for $ \varsigma \in \{ 0, y \} $. By applying the transformation \Eqref{eq:transformation} to \Eqref{eq:BdG}, the BdG Hamiltonian can be brought into the band-space representation. Here, we simply note the following transformation rules that connect the orbital- and band-space representations of the pairing fields:
\be 
[\hat{\Delta}({\bf k})]_{b\kappa,b^{\prime}\kappa^{\prime}} & = & 
\sum_{l,l^{\prime}}\sum_{\mu,\mu^{\prime}}\sum_{\sigma,\sigma^{\prime}}
[\mathcal{U}({\bf k})]_{l\mu\sigma,b\kappa}^{\ast}
[\mathcal{U}(-{\bf k})]_{l^{\prime}\mu^{\prime}\sigma^{\prime},b^{\prime}\kappa^{\prime}}^{\ast}
[\Delta({\bf k})]_{l\mu\sigma,l^{\prime}\mu^{\prime}\sigma^{\prime}}, \\[0.5em]
[\bar{\hat{\Delta}}({\bf k})]_{b\kappa,b^{\prime}\kappa^{\prime}} & = & 
\sum_{l,l^{\prime}}\sum_{\mu,\mu^{\prime}}\sum_{\sigma,\sigma^{\prime}}
[\mathcal{U}(-{\bf k})]_{l\mu\sigma,b\kappa}
[\mathcal{U}({\bf k})]_{l^{\prime}\mu^{\prime}\sigma^{\prime},b^{\prime}\kappa^{\prime}}
[\Delta({\bf k})]_{l\mu\sigma,l^{\prime}\mu^{\prime}\sigma^{\prime}}.
\ee
The band-space pairing fields in turn can be decomposed into pseudospin-singlet and pseudospin-triplet components:
\be 
[\hat{\Delta}({\bf k})]_{b\kappa,b^{\prime}\kappa^{\prime}} = 
\sqrt{2} \sum_{\varsigma} s_{\varsigma}  [\hat{\Delta}_{\varsigma}({\bf k})]_{b,b^{\prime}} [\hat{\Gamma}_{\varsigma}]_{\kappa,\kappa^{\prime}}, \quad
[\bar{\hat{\Delta}}({\bf k})]_{b\kappa,b^{\prime}\kappa^{\prime}} = 
\sqrt{2} \sum_{\varsigma} [\bar{\hat{\Delta}}_{\varsigma}({\bf k})]_{b,b^{\prime}} [\hat{\Gamma}_{\varsigma}]_{\kappa,\kappa^{\prime}},
\ee
where we notationally distinguish BH matrices in pseudospin space by a hat. The pseudospin basis is constructed such that for vanishing SOC it coincides with the physical spin. Correspondingly, in this limit the unitary transformation diagonalizing \Eqref{eq:hamiltonian} factorizes as
$ [\mathcal{U}({\bf k})]_{l\mu\sigma,b\kappa} = [\mathcal{U}({\bf k})]_{l\mu,b^{\prime}}\delta_{\sigma,\kappa} $. For finite SOC, the discussion of inter- and intraband pairing is thus necessarily tied to the pseudospin degree of freedom. 
 
%-------------------------------------------------------------------------------------
\section{Nambu-Gorkov Greens Function}
\label{sec:NG}
%-------------------------------------------------------------------------------------

Having defined the BdG Hamiltonian in \Eqref{eq:BdG}, the corresponding imaginary-time Nambu-Gorkov Green function can be obtained as 
\be 
\mathcal{G}(\mathrm{i}\omega_{n},{\bf k}) = - \int_{0}^{\beta} \! d\tau \mathrm{e}^{\mathrm{i} \omega_{n} \tau} \langle \mathcal{T}_{\tau} \Psi_{{\bf k}}(\tau) \Psi_{{\bf k}}^{\dagger}(0) \rangle_{\scriptsize\mathrm{BCS}},
\ee
where the expectation value is evalueated with respect to a thermal Gibbs state of inverse temperature $\beta = 1/k_{\mathrm{B}}T$ of the BCS Hamiltonian \Eqref{eq:BCS} and $ \omega_{n} = \frac{2\pi}{\beta} (n + 1/2) $, $n \in \mathbbm{Z}$ denotes a fermionic Matsubara frequency. We then decompose the Nambu-Gorkov Green function in the same way as the BdG Hamiltonian to obtain
\be   
\mathcal{G}(\mathrm{i}\omega_{n},{\bf k}) =
\mathcal{P}_{+} \otimes G_{+}(\mathrm{i}\omega_{n},{\bf k})  - 
\mathcal{P}_{-} \otimes  G_{-}(\mathrm{i}\omega_{n},{\bf k})  +
\tau_{+} \otimes F(\mathrm{i}\omega_{n},{\bf k})  + \tau_{-}  \otimes \bar{F}(\mathrm{i}\omega_{n},{\bf k}).
\ee
In the following, we will specialize to purely intraband superconductivity, i.e., we will assume $ [\hat{\Delta}({\bf k})]_{b\kappa,b^{\prime}\kappa^{\prime}} = 0 $ for $ b \neq b^{\prime} $. In order to arrive at the expressions for the particle-hole components of the Nambu-Gorkov Greens function in the orbital representation, we first define a series of auxiliary quantities. We define 
$ E_{b,\kappa}({\bf k}) = \epsilon_{b,\kappa}({\bf k}) - \mu_{0} $, 
$ \hat{\Delta}_{\varsigma}^{b}({\bf k}) =  [\hat{\Delta}_{\varsigma}({\bf k})]_{b,b} $, 
$ \Omega_{0}({\bf k}) = \sum_{\varsigma} \bar{\hat{\Delta}}_{\varsigma}^{b}({\bf k}) \hat{\Delta}_{\varsigma}^{b}({\bf k}) $,
$ \Omega_{\varsigma}(\bf k) = \Omega_{\varsigma}^{(1)}({\bf} k) + \Omega_{\varsigma}^{(2)}({\bf} k)$ (for $ \varsigma \neq 0 $), 
$ \tilde{\Omega}_{\varsigma}(\bf k) = \Omega_{\varsigma}^{(1)}({\bf} k) - \Omega_{\varsigma}^{(2)}({\bf} k)$, where
$ \Omega_{\varsigma}^{(1)}({\bf k}) = \bar{\hat{\Delta}}_{\varsigma}^{b}({\bf k}) \hat{\Delta}_{0}^{b}({\bf k}) + \bar{\hat{\Delta}}_{0}^{b}({\bf k}) \hat{\Delta}_{\varsigma}^{b}({\bf k})$ and 
$ \Omega_{\varsigma}^{(2)}({\bf k}) = \mathrm{i} \sum_{i,j} \epsilon_{\varsigma ij} \bar{\hat{\Delta}}_{i}^{b}({\bf k}) \hat{\Delta}_{j}^{b}({\bf k}) $, as well as $ \mathcal{E}_{b,\kappa}({\bf k}) = \sqrt{E_{b,\kappa}^2({\bf k}) + \Omega_{0}({\bf k}) }$. Equipped with these definitions, we obtain the following general expressions valid for intraband superconductors
\be
\label{eq:Gp} 
[G_{+}(\mathrm{i}\omega_{n},{\bf k})]_{l\mu\sigma,l^{\prime}\mu^{\prime}\sigma^{\prime}} & = &
- \sum_{b,\kappa,\kappa^{\prime}} 
[\mathcal{U}({\bf k})]_{l\mu\sigma,b\kappa}
[\mathcal{U}({\bf k})]_{l^{\prime}\mu^{\prime}\sigma^{\prime},b\kappa^{\prime}}^{\ast}
\frac{\left(\mathrm{i}\omega_{n} + E_{b,\kappa}({\bf k})\right)\left[ \left(\omega_{n}^2 + \mathcal{E}_{b,\kappa}^2({\bf k})\right) \hat{\mathbbm{1}}
+ \sum_{\varsigma}^{\prime} s_{\varsigma} \tilde{\Omega}_{\varsigma}({\bf k}) \hat{\sigma}_{\varsigma}
 \right]_{\kappa,\kappa^{\prime}}
}{\left(\omega_{n}^2 + \mathcal{E}_{b,\kappa}^2({\bf k})\right)^2 - \sum_{\varsigma}^{\prime}\tilde{\Omega}_{\varsigma}^{2}({\bf k})}, \nn \\
\ee
\be
\label{eq:F} 
[F(\mathrm{i}\omega_{n},{\bf k})]_{l\mu\sigma,l^{\prime}\mu^{\prime}\sigma^{\prime}} & = &
\sum_{b,\kappa,\kappa^{\prime}} 
[\mathcal{U}({\bf k})]_{l\mu\sigma,b\kappa}
[\mathcal{U}(-{\bf k})]_{l^{\prime}\mu^{\prime}\sigma^{\prime},b\kappa^{\prime}}
\frac{\left[ \left(\omega_{n}^2 + \mathcal{E}_{b,\kappa}^2({\bf k})\right) \hat{\mathbbm{1}}
+ \sum_{\varsigma}^{\prime} s_{\varsigma} \tilde{\Omega}_{\varsigma}({\bf k}) \hat{\sigma}_{\varsigma}
 \right]
\left[\sum_{\varsigma^{\prime}}\sqrt{2} s_{\varsigma^{\prime}} \hat{\Delta}_{\varsigma^{\prime}}^{b}({\bf k}) \hat{\Gamma}_{\varsigma^{\prime}}\right]_{\kappa,\kappa^{\prime}} }
{\left(\omega_{n}^2 + \mathcal{E}_{b,\kappa}^2({\bf k})\right)^2 - \sum_{\varsigma}^{\prime}\tilde{\Omega}_{\varsigma}^{2}({\bf k})}, \nn \\
\ee
\be 
\label{eq:Fbar}
[\bar{F}(\mathrm{i}\omega_{n},{\bf k})]_{l\mu\sigma,l^{\prime}\mu^{\prime}\sigma^{\prime}} & = &
\sum_{b,\kappa,\kappa^{\prime}} 
[\mathcal{U}(-{\bf k})]_{l\mu\sigma,b\kappa}^{\ast}
[\mathcal{U}({\bf k})]_{l^{\prime}\mu^{\prime}\sigma^{\prime},b\kappa^{\prime}}^{\ast}
\frac{\left[ \left(\omega_{n}^2 + \mathcal{E}_{b,\kappa}^2({\bf k})\right) \hat{\mathbbm{1}}
- \sum_{\varsigma}^{\prime} \Omega_{\varsigma}({\bf k}) \hat{\sigma}_{\varsigma}
 \right]
\left[\sum_{\varsigma^{\prime}}\sqrt{2} \hat{\Delta}_{\varsigma^{\prime}}^{b}({\bf k}) \hat{\Gamma}_{\varsigma^{\prime}}\right]_{\kappa,\kappa^{\prime}} }
{\left(\omega_{n}^2 + \mathcal{E}_{b,\kappa}^2({\bf k})\right)^2 - \sum_{\varsigma}^{\prime}\tilde{\Omega}_{\varsigma}^{2}({\bf k})}, \nn \\
\ee
where $ \sum_{\varsigma} ( \dots )$ runs over $\varsigma = 0,x,y,z$ and $\sum_{\varsigma}^{\prime} ( \dots ) $ is restricted to the pseudospin triplet components $x,y,z$. We also have $ G_{-}(\mathrm{i}\omega_{n},{\bf k}) = [G_{+}(-\mathrm{i}\omega_{n},-{\bf k})]^T $.

Specializing further to a pseudospin singlet superconductor, i.e., $\hat{\Delta}_{0}^{b}({\bf k}) \neq 0$, while $ \hat{\Delta}_{x}^{b}({\bf k}) = \hat{\Delta}_{y}^{b}({\bf k}) = \hat{\Delta}_{z}^{b}({\bf k}) = 0 $ throughout the Brillouin zone, the expressions Eqs.~(\ref{eq:Gp})-(\ref{eq:Fbar}) simplify to
\be 
\label{eq:Gp_singlet}
[G_{+}(\mathrm{i}\omega_{n},{\bf k})]_{l\mu\sigma,l^{\prime}\mu^{\prime}\sigma^{\prime}} 
& = & 
- \sum_{b,\kappa,\kappa^{\prime}} 
[\mathcal{U}({\bf k})]_{l\mu\sigma,b\kappa}
[\mathcal{U}({\bf k})]_{l^{\prime}\mu^{\prime}\sigma^{\prime},b\kappa^{\prime}}^{\ast}
\frac{\mathrm{i}\omega_{n} + E_{b,\kappa}}{\omega_{n}^2 + \mathcal{E}_{b,\kappa}^2({\bf k})}[\hat{\mathbbm{1}}]_{\kappa,\kappa^{\prime}}, \\[0.5em]
\label{eq:F_singlet}
[F(\mathrm{i}\omega_{n},{\bf k})]_{l\mu\sigma,l^{\prime}\mu^{\prime}\sigma^{\prime}} 
& = &
- \sqrt{2}\sum_{b,\kappa,\kappa^{\prime}} 
[\mathcal{U}({\bf k})]_{l\mu\sigma,b\kappa}
[\mathcal{U}(-{\bf k})]_{l^{\prime}\mu^{\prime}\sigma^{\prime},b\kappa^{\prime}}
\frac{\hat{\Delta}_{0}^{b}({\bf k})}{\omega_{n}^2 + \mathcal{E}_{b,\kappa}^{2}({\bf k})}
[\hat{\Gamma}_{0}]_{\kappa,\kappa^{\prime}}, \\[0.5em]
\label{eq:Fbar_singlet}
[\bar{F}(\mathrm{i}\omega_{n},{\bf k})]_{l\mu\sigma,l^{\prime}\mu^{\prime}\sigma^{\prime}} & = &
+\sqrt{2}\sum_{b,\kappa,\kappa^{\prime}} 
[\mathcal{U}(-{\bf k})]_{l\mu\sigma,b\kappa}^{\ast}
[\mathcal{U}({\bf k})]_{l^{\prime}\mu^{\prime}\sigma^{\prime},b\kappa^{\prime}}^{\ast}
\frac{\bar{\hat{\Delta}}_{0}^{b}({\bf k})}{\omega_{n}^2 + \mathcal{E}_{b,\kappa}^{2}({\bf k})}
[\hat{\Gamma}_{0}]_{\kappa,\kappa^{\prime}}.
\ee
%

%-------------------------------------------------------------------------------------
\section{Bare Particle-Hole Correlation Function of a Spin-Orbit Coupled Superconductor}
\label{sec:bare}
%-------------------------------------------------------------------------------------

Here we collect results on the generalized, connected correlation function (also referred to as irreducible bubble) evaluated for a BCS Hamiltonian with SOC for an intraband, pseudospin singlet superconductor. We define 
\be
\label{eq:chi0}
[\chi_{0}(\mathrm{i}\omega_n,{\bf q})]^{l_1\mu_1\sigma_1;l_2\mu_2\sigma_2}_{l_3\mu_3\sigma_3; l_4\mu_4\sigma_4} =
\frac{1}{\mathcal{N}}\int_{0}^{\beta} \! d\tau \, \mathrm{e}^{\mathrm{i}\omega_n \tau}\sum_{{\bf k},{\bf k}^{\prime}}
\langle \mathcal{T}_{\tau} 
c_{{\bf k} - {\bf q}l_1\mu_1\sigma_1}^{\dagger}(\tau) 
c_{{\bf k}l_2\mu_2\sigma_2}(\tau)
c_{{\bf k}^{\prime} + {\bf q}^{\prime}l_3\mu_3\sigma_3}^{\dagger}(0) 
c_{{\bf k}^{\prime}l_4\mu_4\sigma_4}(0)
\rangle_{\scriptsize{c,\mathrm{BCS}}},
\ee
with $\omega_n$ now denoting a bosonic Matsubara frequency $\frac{2\pi}{\beta} n$, $ n \in \mathbbm{Z}$. Applying Wick's theorem, we arrive at the following representation of the irreducible bubble in terms of the components of the Nambu-Gorkov Greens function:
\be 
[\chi_{0}(\mathrm{i}\omega_n,{\bf q})]^{l_1\mu_1\sigma_1;l_2\mu_2\sigma_2}_{l_3\mu_3\sigma_3; l_4\mu_4\sigma_4} & = & 
- \frac{1}{\beta \mathcal{N}} \sum_{k \in \mathbbm{Z}, {\bf k}}
[G_{+}(\mathrm{i}\nu_{k},{\bf k})]_{l_{2}\mu_{2}\sigma_{2},l_{3}\mu_{3}\sigma_{3}}
[G_{+}(\mathrm{i}\nu_{k}  - \mathrm{i}\omega_{n},{\bf k} - {\bf q})]_{l_{4}\mu_{4}\sigma_{4},l_{1}\mu_{1}\sigma_{1}} \nn \\
& & - \frac{1}{\beta \mathcal{N}} \sum_{k \in \mathbbm{Z}, {\bf k}}
[\bar{F}(\mathrm{i}\nu_{k},{\bf k})]_{l_{1}\mu_{1}\sigma_{1},l_{3}\mu_{3}\sigma_{3}}
[F(\mathrm{i}\nu_{k}  - \mathrm{i}\omega_{n},{\bf k} - {\bf q})]_{l_{4}\mu_{4}\sigma_{4},l_{2}\mu_{2}\sigma_{2}}.
\ee
Using Eqs.~(\ref{eq:Gp_singlet})-(\ref{eq:Fbar_singlet}) and performing Matsubara summations, 
we eventually arrive at 
\be 
[\chi_{0}(\mathrm{i}\omega_n,{\bf q})]^{l_1\mu_1\sigma_1;l_2\mu_2\sigma_2}_{l_3\mu_3\sigma_3; l_4\mu_4\sigma_4} & = & 
-\frac{1}{\mathcal{N}}\sum_{{\bf k}}\sum_{b,b^{\prime}} 
[\mathcal{M}_{b,b^{\prime}}^{GG}({\bf k},{\bf q})]^{l_{1}\mu_{1}\sigma_{1};l_{2}\mu_{2}\sigma_{2}}_{l_{3}\mu_{3}\sigma_{3};l_{4}\mu_{4}\sigma_{4}}
\mathcal{L}^{GG}(\mathrm{i}\omega_{n},E_{b}({\bf k}),E_{b^{\prime}}({\bf k}-{\bf q}),
\mathcal{E}_{b}({\bf k}),\mathcal{E}_{b^{\prime}}({\bf k}-{\bf q})) \nn \\
& & -\frac{1}{\mathcal{N}}\sum_{{\bf k}}\sum_{b,b^{\prime}} 
[\mathcal{M}_{b,b^{\prime}}^{\bar{F}F}({\bf k},{\bf q})]^{l_{1}\mu_{1}\sigma_{1};l_{2}\mu_{2}\sigma_{2}}_{l_{3}\mu_{3}\sigma_{3};l_{4}\mu_{4}\sigma_{4}}
\mathcal{L}^{\bar{F}F}(\mathrm{i}\omega_{n},\mathcal{E}_{b}({\bf k}),\mathcal{E}_{b^{\prime}}({\bf k}-{\bf q}),\bar{\hat{\Delta}}_{0}^{b}({\bf k}),\hat{\Delta}_{0}^{b^{\prime}}({\bf k} - {\bf q})),
\ee
where we dropped the pseudospin index $\kappa$ on energy arguments, as we assume time-reversal-invariance and inversion-symmetry, implying $ E_{b,+}({\bf k}) =  E_{b,-}({\bf k})$ and $ \mathcal{E}_{b,+}({\bf k}) =  \mathcal{E}_{b,-}({\bf k})$. The coefficients and Lindhard factors are defined as
\be 
[\mathcal{M}_{b,b^{\prime}}^{GG}({\bf k},{\bf q})]^{l_{1}\mu_{1}\sigma_{1};l_{2}\mu_{2}\sigma_{2}}_{l_{3}\mu_{3}\sigma_{3};l_{4}\mu_{4}\sigma_{4}} & =  & \!\!\!\!\!\!\!\!
\sum_{\kappa_{1},\kappa_{2},\kappa_{3},\kappa_{4}}  \!\!\!\!\!\!
[\mathcal{U}({\bf k}-{\bf q})]_{l_{1}\mu_{1}\sigma_{1},b^{\prime}\kappa_{1}}^{\ast}
[\mathcal{U}({\bf k})]_{l_{2}\mu_{2}\sigma_{2},b\kappa_{2}}
[\mathcal{U}({\bf k})]_{l_{3}\mu_{3}\sigma_{3},b\kappa_{3}}^{\ast}
[\mathcal{U}({\bf k}-{\bf q})]_{l_{4}\mu_{4}\sigma_{4},b^{\prime}\kappa_{4}}
[\hat{\mathbbm{1}}]_{\kappa_{1},\kappa_{4}}[\hat{\mathbbm{1}}]_{\kappa_{2},\kappa_{3}}, \nn
\ee
\be
[\mathcal{M}_{b,b^{\prime}}^{\bar{F}F}({\bf k},{\bf q})]^{l_{1}\mu_{1}\sigma_{1};l_{2}\mu_{2}\sigma_{2}}_{l_{3}\mu_{3}\sigma_{3};l_{4}\mu_{4}\sigma_{4}} & =  & \!\!\!\!\!\!\!\!
\sum_{\kappa_{1},\kappa_{2},\kappa_{3},\kappa_{4}} \!\!\!\!\!\!
[\mathcal{U}(-{\bf k})]_{l_{1}\mu_{1}\sigma_{1},b\kappa_{1}}^{\ast}
[\mathcal{U}(-{\bf k}+{\bf q})]_{l_{2}\mu_{2}\sigma_{2},b^{\prime}\kappa_{2}}
[\mathcal{U}({\bf k})]_{l_{3}\mu_{3}\sigma_{3},b\kappa_{3}}^{\ast}
[\mathcal{U}({\bf k}-{\bf q})]_{l_{4}\mu_{4}\sigma_{4},b^{\prime}\kappa_{4}}
s_{0}[\hat{\Gamma}_{0}]_{\kappa_{1},\kappa_{3}}[\hat{\Gamma}_{0}]_{\kappa_{4},\kappa_{2}}\nn 
\ee
and
\be 
\mathcal{L}^{GG}(\Omega,E_{1},E_{2},\mathcal{E}_{1},\mathcal{E}_{2}) & = &
\frac{1}{4}
\frac{\left(E_{1} + \mathcal{E}_{1}\right)\left(E_{2} + \mathcal{E}_{2}\right)}{\mathcal{E}_{1}\mathcal{E}_{2}}
\frac{n_{\mathrm{F}}(\mathcal{E}_{2}) - n_{\mathrm{F}}(\mathcal{E}_{1})}
{\Omega + \mathcal{E}_{2} - \mathcal{E}_{1}} + \frac{1}{4}
\frac{\left(E_{1} - \mathcal{E}_{1}\right)\left(E_{2} + \mathcal{E}_{2}\right)}{\mathcal{E}_{1}\mathcal{E}_{2}}
\frac{1 - n_{\mathrm{F}}(\mathcal{E}_{2}) - n_{\mathrm{F}}(\mathcal{E}_{1})}
{\Omega + \mathcal{E}_{2} + \mathcal{E}_{1}} + \nn \\
& & \frac{1}{4}
\frac{\left(E_{1} + \mathcal{E}_{1}\right)\left(E_{2} - \mathcal{E}_{2}\right)}{\mathcal{E}_{1}\mathcal{E}_{2}}
\frac{n_{\mathrm{F}}(\mathcal{E}_{2}) - n_{\mathrm{F}}(\mathcal{E}_{1}) - 1}
{\Omega - \mathcal{E}_{2} - \mathcal{E}_{1}} + \frac{1}{4}
\frac{\left(E_{1} - \mathcal{E}_{1}\right)\left(E_{2} - \mathcal{E}_{2}\right)}{\mathcal{E}_{1}\mathcal{E}_{2}}
\frac{- n_{\mathrm{F}}(\mathcal{E}_{2}) + n_{\mathrm{F}}(\mathcal{E}_{1}) }
{\Omega - \mathcal{E}_{2} + \mathcal{E}_{1}}, \nn
\ee
\be 
\mathcal{L}^{\bar{F}F}(\Omega,\mathcal{E}_{1},\mathcal{E}_{2},\Delta_{1},\Delta_{2}) & = &
\frac{1}{4}
\frac{\Delta_{1}\Delta_{2}}{\mathcal{E}_{1}\mathcal{E}_{2}} \left(
\frac{n_{\mathrm{F}}(\mathcal{E}_{2}) - n_{\mathrm{F}}(\mathcal{E}_{1})}
{\Omega + \mathcal{E}_{2} - \mathcal{E}_{1}} + 
\frac{1 - n_{\mathrm{F}}(\mathcal{E}_{2}) - n_{\mathrm{F}}(\mathcal{E}_{1})}
{\Omega + \mathcal{E}_{2} + \mathcal{E}_{1}} + 
\frac{n_{\mathrm{F}}(\mathcal{E}_{2}) - n_{\mathrm{F}}(\mathcal{E}_{1}) - 1}
{\Omega - \mathcal{E}_{2} - \mathcal{E}_{1}} + 
\frac{- n_{\mathrm{F}}(\mathcal{E}_{2}) + n_{\mathrm{F}}(\mathcal{E}_{1}) }
{\Omega - \mathcal{E}_{2} + \mathcal{E}_{1}} \right). \nn
\ee
Above, we also introduced the Fermi-Dirac distribution function $n_{\mathrm{F}}(\mathcal{E}) = 1/(\exp(\beta \mathcal{E}) + 1)$. 

In view of the SU(2) `gauge freedom' related to the degenarcy of eigenstates, which is adressed in detail in Ref.~\onlinecite{SMscherer2019}, it is important to note, that for a fixed Wannier gauge, the irreducible bubble can indeed be shown to be gauge invariant, as long as each quadruplet of states at ${\bf k}$ and $ - {\bf k} $ is constructed from a single state through time-reversal, inversion, and the combined application of time-reversal and inversion.

%-------------------------------------------------------------------------------------
\section{RPA Correlation Function of a Spin-Orbit Coupled Superconductor}
\label{sec:RPA}
%-------------------------------------------------------------------------------------

Following Refs.~\onlinecite{SMscherer2016} and \onlinecite{SMscherer2018}, where we developed the RPA in the absence of spin SU(2) symmetry (either due to non-collinear magnetic order or SOC), we briefly describe the RPA formalism required to analyze the magnetic anisotropy of the neutron resonance in the presence of SOC. The goal is to compute the connected imaginary-time spin-spin correlation function (where  $i,j$ refer to the spatial directions $x,y,z$)
\be
\chi^{ij}(\mathrm{i}\omega_n,{\bf q}) = g^2\int_{0}^{\beta} \! d\tau \,
\mathrm{e}^{\mathrm{i}\omega_n \tau}
\langle \mathcal{T}_{\tau} S^{i}_{{\bf q}}(\tau) S^{j}_{-{\bf q}}(0)\rangle_{\scriptsize{c,\mathrm{BCS}+\mathrm{RPA}}},
\ee
with the Fourier transformed electron spin operator for the 2-Fe unit cell given as
\be
S^{i}_{{\bf q}}(\tau) = \frac{1}{\sqrt{\mathcal{N}}}\sum_{{\bf k},l,\mu,\sigma,\sigma^{\prime}} c_{{\bf k} - {\bf q}l\mu\sigma}^{\dagger}(\tau) \frac{\sigma_{\sigma\sigma^{\prime}}^{i}}{2}c_{{\bf k}l\mu\sigma^{\prime}}(\tau).
\ee
We note that we typically specify the transfer momentum $ {\bf q} $ with respect to the coordinate system of the 1-Fe Brillouin zone. It is then understood that `$ {\bf k} - {\bf q} $' refers to subtraction of the two vectors in a common coordinate system. Here $\mathcal{T}_{\tau}$ denotes the time-ordering operator with respect to the imaginary-time variable $\tau \in [0,\beta)$, with $\beta$ the inverse temperature and $\sigma_{\sigma\sigma^{\prime}}^{i}$ the $i$-th Pauli matrix. From the imaginary part of $\chi^{ij}(\omega,{\bf q})$ (after having analytically continued from imaginary to real frequencies), we can extract the spectrum of spin excitations that are probed by neutron scattering. 

To ease notation we introduce a combined index $X \equiv (l,\mu,\sigma)$ by collecting sublattice, orbital and spin indices. In the absence of interactions, the correlation function $[\chi(\mathrm{i}\omega_n,{\bf q}) ]^{X_1;X_2}_{X_3;X_4}\equiv
[\chi(\mathrm{i}\omega_n,{\bf q})]^{l_1\mu_1\sigma_1;l_2\mu_2\sigma_2}_{l_3\mu_3\sigma_3; l_4\mu_4\sigma_4} $ reduces to
\be 
[\chi_{0}(\mathrm{i}\omega_n,{\bf q})]^{X_1;X_2}_{X_3;X_4} =
\frac{1}{\mathcal{N}}\int_{0}^{\beta} \! d\tau \, \mathrm{e}^{\mathrm{i}\omega_n \tau}\sum_{{\bf k},{\bf k}^{\prime}}
\langle \mathcal{T}_{\tau} 
c_{{\bf k} - {\bf q}X_1}^{\dagger}(\tau) 
c_{{\bf k}X_2}(\tau)
c_{{\bf k}^{\prime} + {\bf q}^{\prime}X_3}^{\dagger}(0) 
c_{{\bf k}^{\prime}X_4}(0)
\rangle_{\scriptsize{c, \mathrm{BCS}}},
\ee
which was discussed extensively in Sec.~\ref{sec:bare}. Ignoring anomalous vertices, the RPA equation for the generalized particle-hole correlation function reads as (see Refs.~\onlinecite{SMscherer2016,SMscherer2018})
\be 
\label{eq:RPAequation}
[\chi(\mathrm{i}\omega_n,{\bf q}) ]^{X_1;X_2}_{X_3; X_4}=
[\chi_{0}(\mathrm{i}\omega_n,{\bf q})]^{X_1;X_2}_{X_3;X_4} +
[\chi_{0}(\mathrm{i}\omega_n,{\bf q})]^{X_1;X_2}_{Y_1;Y_2}
[U]^{Y_1;Y_2}_{Y_3;Y_4}
[\chi(\mathrm{i}\omega_n,{\bf q})]^{Y_3;Y_4}_{X_3;X_4}.
\ee
Repeated indices are summed over in \Eqref{eq:RPAequation}. The bare fluctuation vertex $[U]^{X_1;X_2}_{X_3;X_4} \equiv [U]^{l_1\mu_1\tau_{1};l_2\mu_2\tau_{2}}_{l_3\mu_3\tau_{3};l_4\mu_4\tau_{4}}$ originates from the Hubbard-Hund interaction and describes how electrons scatter off a collective excitation in the particle-hole channel. The electronic Hubbard-Hund interaction Hamiltonian (with normal ordering implied) can be compactly written as
\be
H_{\mathrm{int}} =
-\frac{1}{2}\sum_{i} \sum_{\{l_{n}\},\{\mu_j \},\{ \sigma_k \}} 
[U]^{l_{1}\mu_{1}\sigma_{1};l_{2}\mu_{2}\sigma_{2}}_{l_{3}\mu_{3}\sigma_{3};l_{4}\mu_{4}\sigma_{4}}
c_{l_{1}i\mu_{1}\sigma_{1}}^{\dagger}
c_{l_{2}i\mu_{2}\sigma_{2}}
c_{l_{3}i\mu_{3}\sigma_{3}}^{\dagger}
c_{l_{4}i\mu_{4}\sigma_{4}}.
\ee
The bare interaction vertex $[U]^{\mu_{1}\sigma_{1};\mu_{2}\sigma_{2}}_{\mu_{3}\sigma_{3};\mu_{4}\sigma_{4}}$ defined above can be decomposed into charge and spin vertices in the particle-hole channel as
\be
[U]^{l_{1}\mu_{1}\sigma_{1};l_{2}\mu_{2}\sigma_{2}}_{l_{3}\mu_{3}\sigma_{3};l_{4}\mu_{4}\sigma_{4}}
= - \frac{1}{2} \delta_{l_{1},l_{2}}\delta_{l_{2},l_{3}}\delta_{l_{3},l_{4}}
\left( 
[U_{\mathrm{c}}]^{\mu_{1}\mu_{2}}_{\mu_{3}\mu_{4}} \,
\delta_{\sigma_{1}\sigma_{2}}\delta_{\sigma_{3}\sigma_{4}}  - 
[U_{\mathrm{s}}]^{\mu_{1}\mu_{2}}_{\mu_{3}\mu_{4}} \,
{\boldsymbol \sigma}_{\sigma_{1}\sigma_{2}} \cdot {\boldsymbol \sigma}_{\sigma_{3}\sigma_{4}}
\right),
\ee
which in turn are defined by
\be
[U_{\mathrm{s}}]^{\mu\mu}_{\mu\mu} = U,
\quad
[U_{\mathrm{s}}]^{\nu\mu}_{\mu\nu} = U^{\prime},
\quad
[U_{\mathrm{s}}]^{\nu\nu}_{\mu\mu} = J,
\quad
[U_{\mathrm{s}}]^{\mu\nu}_{\mu\nu} = J^{\prime},
\quad\text{with}\,\mu \neq \nu,
\ee
and
\be
[U_{\mathrm{c}}]^{\mu\mu}_{\mu\mu} = U,
\quad
[U_{\mathrm{c}}]^{\nu\mu}_{\mu\nu} = 2J - U^{\prime},
\quad
[U_{\mathrm{c}}]^{\nu\nu}_{\mu\mu} = 2U^{\prime}-J,
\quad
[U_{\mathrm{c}}]^{\mu\nu}_{\mu\nu} = J^{\prime},
\quad\text{with}\,\mu \neq \nu,
\ee
and zero otherwise. The onsite interaction is parametrized by an intraorbital Hubbard-$U$, an interorbital coupling $U^{\prime}$, Hund's coupling $J$ and pair hopping $J^{\prime}$. 

Solving \Eqref{eq:RPAequation}, we obtain the RPA approximation for the particle-hole correlation function. The RPA approximation for the spin susceptibility tensor $ \chi^{ij}(\omega,{\bf q}) $ can be obtained by forming the appropriate linear combinations of components of the particle-hole correlation function and performing analytic continuation $ \mathrm{i}\omega_{n} \to \omega + \mathrm{i}\eta $, $\eta > 0$:
\be
\label{eq:susc_final}
\chi^{ij}(\omega,{\bf q}) & = & 
g^{2}\sum_{l,l^{\prime}} \sum_{\mu,\nu} \sum_{\sigma_{1},\dots,\sigma_{4}}
\frac{\sigma_{\sigma_{1}\sigma_{2}}^{i}}{2} \frac{\sigma_{\sigma_{3}\sigma_{4}}^{j}}{2}
[\chi(\mathrm{i}\omega_n\to\omega + \mathrm{i}\eta,{\bf q})]^{l\mu\sigma_{1};l\mu\sigma_{2}}_{l^{\prime}\nu\sigma_3; l^{\prime}\nu\sigma_4}.
\ee
%

%-------------------------------------------------------------------------------------

%-------------------------------------------------------------------------------------

\end{widetext}

\end{document}